\documentclass[twocolumn,aps,prl,showpacs]{revtex4-2}
\usepackage{amssymb,amsmath,amsfonts,epsfig,bm}

\usepackage{color}
\usepackage[citecolor=blue,colorlinks=true,linkcolor=blue,urlcolor=blue]{hyperref}
\usepackage{orcidlink}
\begin{document}

\title{Spin-orientation-resolved attosecond chronoscopy in strong field ionization}

\author{Mingqing Liu \orcidlink{0000-0002-6098-0798}}
\affiliation{Institute of Quantum Precision Measurement, College of Physics and Optoelectronic Engineering, Shenzhen University, Shenzhen 518060, China}

\author{Wei-Chao Jiang \orcidlink{0000-0003-4372-0393}}
\email[]{jiang.wei.chao@szu.edu.cn}
\affiliation{Institute of Quantum Precision Measurement, College of Physics and Optoelectronic Engineering, Shenzhen University, Shenzhen 518060, China}

\date{\today}

\begin{abstract}
Attosecond chronoscopy represents a major breakthrough in the study of ultrafast phenomena and has the potential to revolutionize our understanding of the fundamental physics of matter.
We theoretically investigate the spin-orientation-resolved attosecond chronoscopy
for the first time by the circular RABBIT technique for Kr atoms.
Due to the spin-orbit interaction and the sensitivity of ionization in circularly polarized fields to
the sense of electron rotation in the initial state, the spin-resolved ionization rates of photoelectrons emitted from $^2P_{1/2}$ and $^2P_{3/2}$ channels can be expressed via $m$-resolved ionization rates of initial state, where $m$ is the orbital magnetic quantum number.
We demonstrate that the yields difference between spin-up and spin-down photoelectrons
from each channel are closely associated with the different behaviors of corresponding Wigner time delay.
We find that the Wigner time delay between spin-up and spin-down photoelectrons in the polarization plane can reach several tens of attoseconds in the co-rotating geometry, but a few attoseconds in the counter-rotating geometry.
Our approach opens up a new avenue for probing the spin-dependent behavior of Wigner time delay, and lays the foundation for spin-orientation-resolved attosecond chronoscopy, which can be verified by the current experimental techniques.
\end{abstract}

\maketitle
%%%%%%%%%%%%%%%%%%%%%%%%%%%%%%%%%%%%%%%%%%%%%%%%%%%%%%%%%%%%%%%%%%%%%%%%%%%
%\section{INTRODUCTION}
Observing and controlling the electronic dynamics in real time, referred to as attosecond chronoscopy \cite{Pazourek2015}, always are a dream dedicated to achieving in attosecond science \cite{Corkum2007,Krausz2009,Peng2015}, where the challenge is the requirement of attosecond-scale (1 as $=10^{-18}$ s) time-resolution spectroscopy. It can be accessible after the advent of broadband coherent extreme ultraviolet (XUV) sources generated from intense infrared (IR) pulses through high-order harmonic generation  \cite{Drescher2001,Hentschel2001,Paul2001}.
The photoemission time delay in one-photon ionization, called Wigner time, can be expressed as the energy derivative of the scattering phase of a photoelectron wave packet \cite{Wigner1955,Smith1960}.
In order to measure precisely this scattering phase on the natural timescale, an attosecond interferometry named RABBIT (reconstruction of attosecond beating by interference of two-photon transitions) technique, which is based on a XUV-attosecond pulse train (APT) as a pump with a phase-controlled near-infrared (NIR) pulse as a probe, is complemented in atoms \cite{Paul2001,Muller2002,Swoboda2010,Klunder2011,Cirelli2018,Zitnik2022,Autuori2022}, molecules \cite{Haessler2009,Rist2021,Gong2022}, and solid targets \cite{Locher2015}.
The amplitude of the sidebands (SBs) yield can be written as $A_{\mathrm{SB}}\propto \cos(2\omega\tau-\phi_\mathrm{RABBIT})$ ,
where the RABBIT phase $\phi_\mathrm{RABBIT}$ encodes atomic photoionization time delay \cite{Klunder2011,Fuchs2021}.

In conventional RABBIT, a linearly polarized XUV-APT with odd harmonics ionize the target. Nowadays, the circularly polarized XUV harmonics are available experimentally \cite{Kfir2014,Fan2015,Barreau2018,Donsa2019}.
When an atom with the $p$ outer shell is ionized via circularly polarized XUV harmonics, the spin polarization arises from spin-orbit interaction \cite{Barth2013,Hartung2016}, which
splits the ionic states with respect to the total angular momentum of the core $J = 1/2$ and $J = 3/2$,
providing two ionization channels with slightly different
ionization potentials.
It has been demonstrated that these ionization
channels can be well distinguished in the above-threshold ionization (ATI) spectra for ultraviolet laser pulses \cite{Nakano2017,Liu2018,Trabert2018,Han2020,Ge2021}. Additionally,
the total ionization rates of spin-up and spin-down photoelectrons can be expressed as a superposition of ionization rates of electrons for different initial states with all possible orbital magnetic quantum numbers $m$.
However, the experimental, even theoretical, investigates on spin-orientation-resolved time delay of photoelectrons from each ionization channel by circular RABBIT are scarce so far.

Recently, Han \emph{et al.} \cite{Han2022} used circular-RABBIT attosecond metrology to manipulate and probe the chiral dynamics of atomic, specifically exploring electron vortices \cite{Djiokap2015,Djiokap2016,Yuan2016,Pengel2017,Li2018,Geng2021} from continuum to continuum states.
This innovative technique, dubbed as attosecond circular-dichroism
chronoscopy, allows the clocking of continuum-continuum (cc) transitions and brings the dream of time-resolved quantum physics a little closer \cite{Djiokap2022}.
Besides, the vortex-shaped photoelectron from the two ionization channels can be well dislocated in momentum space, leading to the spin polarization exceeding $50\%$ \cite{Hu2023}.
However, it is still unclear whether the spin-up electrons emitted from one of channels are delayed or advanced, or even instantaneous in comparison to the spin-down photoelectrons from the same channel.

\begin{figure}[tb]
  \centering
  \includegraphics[scale=0.56]{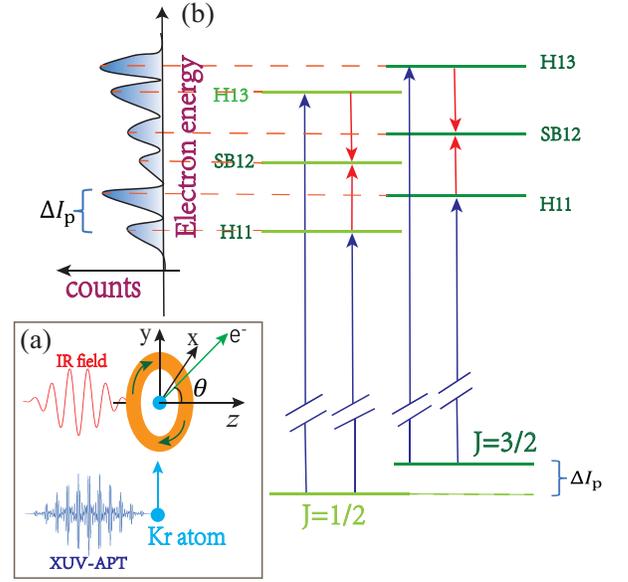}
  \caption{(a) Schematic of circular RABBIT protocol. The polarization plane of laser pulse is the $x-y$ plane and the
light propagation direction defines the $z$ axis, where $\theta$ is the photoelectron emission angle with respect to the light propagation direction. (b) Sketch map of energy spectrum of spin-polarized photoelectrons. Blue arrows indicate photoionization induced by the XUV (only 11th and 13th harmonics are shown here), and red arrows indicate continuum-continuum transitions induced by the IR (absorption or emission of one IR photon). The ground state ($^2P$, $J=3/2$) and first excited state ($^2P$, $J=1/2$) of Kr$^+$ differ by $\Delta I_p=0.66$ eV in ionization potential, therefore, the photoelectron yields show the two combs of peaks spaced by the photon energy with a relative offset of 0.66 eV.}\label{fig1}
\end{figure}

In this Letter, we investigate spin-orientation-resolved attosecond chronoscopy by employing the circular RABBIT technique.
Specifically, the left circularly polarized XUV-APT ionizes the outer shell of Kr atom, creating a spin-polarized electron-vortex continuum state with a well-defined helicity, which are probed by a time-delayed synchronized co-rotation or counter-rotating IR laser pulse IR field, as shown in Fig.~\ref{fig1}(a).
Our results reveal that the SB yield of spin-up photoelectrons from $^2P_{1/2}$ ($^2P_{3/2}$) channel is remarkably suppressed (enhanced) as compared to the spin-down photoelectrons from the same channel in the co-rotating geometry.
However, the SB yields between the spin-up and spin-down photoelectrons are comparable regardless of the ionization channels in the counter-rotating geometry.
We find that, in the co-rotating case, the Wigner time delays
between the spin-up and spin-down photoelectrons emitted parallel to the light-polarization plane differ significantly, with differences of 67 attoseconds observed for the $^2P_{1/2}$ channel and -22 attoseconds for the $^2P_{3/2}$ channel. In contrast, in the counter-rotating configuration, this time delay is only a few attoseconds.
Our study presents an experimentally viable circular RABBIT scheme for achieving spin-orientation-resolved attosecond chronoscopy.

%\section{Theoretical framework}
We numerically solve the three-dimensional time-dependent
Schr\"{o}dinger equation (TDSE) within the single-active electron
approximation. In the velocity gauge, the TDSE
reads (atomic units are used throughout unless stated otherwise)
\begin{equation}\label{tdse}
  i\frac{\partial}{\partial t}\psi(\mathbf{r},t)=\left[-\frac{\nabla^2}{2}+V(\mathbf{r})-i\mathbf{A}(t)\cdot\nabla\right]\psi(\mathbf{r},t)
\end{equation}
where $\psi(\mathbf{r},t)$ is time-dependent electron wave function, and $\mathbf{A}(t)$ is the vector potential.
We use a parametric model potential \cite{Cloux2015},
\begin{equation}\label{pot}
  V(r)=-\frac{1+A\exp(-Br)+(N-A)\exp(-Cr)}{r},
\end{equation}
to represent the electron-core interaction for the Krypton atom.
In this model potential, $N = 35$ is the number of core electrons, and ${A,B,C}$ are positive parameters, which are optimized to reproduce as accurately
as possible the valence state energies of Kr atom.
To reproduce the ionization potentials of two ionization channels due to the spin-orbit interaction, we adjust the parameters $A=6.42$, $B=0.905$, and $C=4.1996$ to match the ionization potentials $I_p^{P_{3/2}}=0.5145$ a.u. (14.0 eV), or $C=4.1174$ for $I_p^{P_{1/2}}=0.5389$ a.u. (14.66 eV).
The initial magnetic quantum number $m$ was tuned to be
1, 0 and $-1$ without changing the ionization potential, corresponding to three degenerate orbitals $p_{+}$ (left helicity), $p_0$ and $p_{-}$ (right helicity).

The vector potential of left circularly polarized XUV-APT can be expressed as
\begin{align}\label{xuv}
  \mathbf{A}_{\mathrm{XUV}}(t)=&\sum_{n=11,13,15,17}\frac{\sqrt{I^0_{\mathrm{XUV}}}}{\omega_{\mathrm{XUV}}}
\cos^2(\omega t/2n_c)\nonumber \\
   &\times [\sin(n\omega t)\mathbf{x}+\cos(n\omega t)\mathbf{y}],
\end{align}
and that of the IR field is
\begin{align}\label{ir}
  \mathbf{A}_{\mathrm{IR}}(t)=&\frac{\sqrt{I^0_{\mathrm{IR}}}}{\omega}
\cos^2(\omega t/2n_c)\nonumber \\
   &\times \{\sin[\omega (t+\tau)]\mathbf{x}+\xi\cos[\omega (t+\tau)]\mathbf{y}\},
\end{align}
with amplitudes $I^0_{\mathrm{XUV}}=1\times 10^{11}$ W/cm$^2$,
$I^0_{\mathrm{IR}}=1\times 10^{12}$ W/cm$^2$, and the photon energy of IR field is $\omega=0.057$ a.u. (800 nm). the ellipticity $\xi=1$ and $-1$ correspond to the co-rotating and counter-rotating geometry between XUV-APT and IR pulse, respectively.
The pulse duration amounts to $n_c=6$ optical cycles, and the XUV-IR delay $\tau$ was uniformly sampled by 24 points in one IR cycle.
The momentum box was set from 0.01 a.u. to 1.1 a.u. with 400 bins, and there are 93 uniform bins for $\phi=\arctan(p_y/p_x)$ in the $2\pi$ range and 37 uniform bins for $\theta=\arccos(p_z/p_{\mathrm{total}})$ in the $\pi$ range.

To efficiently propagate the wave function in time, we employ the split-Lanczos propagator \cite{Jiang2017,Jiang2020,Wang2020,Jiang2021,Jiang2022,Liu2022}. Our approach uniquely expresses the wave function as a sum of spherical harmonic functions, while employing the finite-element discrete variable representation (FE-DVR) method \cite{Rescigno2000,Schneider2005,Rayson2007,Geng2022} for discretizing the radial part of the wave function. To circumvent the need for an excessively large radial box, we adopt the wave-splitting technique \cite{Tong2006}.
The maximum angular momentum included is $\ell_{\mathrm{max}}$ = 15, which fully covers all ionized electronic partial waves.
The inner radial box size $R_c=360$ a.u. with maximal box size
$R_c=600$ a.u., and the time step of $\Delta t=0.01$ a.u. are
sufficient to ensure the calculation convergence.

To include the spin-orbit coupling, we superpose the ionization contribution
from atomic orbitals with different magnetic quantum numbers. The relative weight is determined by the Clebsch-Gordan coefficients. According to the relation between the spin and orbital angular momenta, one can infer the spin of photoelectrons. The rates of spin-up and spin-down photoelectrons are given by \cite{Barth2013,milo2016,Liu2018,Hu2023}
\begin{align}\label{up}
  w_{\uparrow,\downarrow}(\mathbf{p})=&\frac{1}{3}w^{p_{0}}(\mathbf{p},I_p^{P_{\frac{1}{2}}})+\frac{2}{3} w^{p_{0}}(\mathbf{p},I_p^{P_{\frac{3}{2}}})\nonumber \\
  &+w^{p_{\pm}}(\mathbf{p},I_p^{P_{\frac{3}{2}}})+\frac{2}{3}w^{p_{\mp}}(\mathbf{p},I_p^{P_{\frac{1}{2}}})\nonumber\\
  &+\frac{1}{3} w^{p_{\mp}}(\mathbf{p},I_p^{P_{\frac{3}{2}}}).
\end{align}
The momentum-resolved total spin polarization $S(\mathbf{p})$ is proportional to the
difference in the total ionization rates for the photoelectrons
with spin-up $w_{\uparrow}(\mathbf{p})$ and spin-down $w_{\downarrow}(\mathbf{p})$: $
  S(\mathbf{p})=\frac{w_{\uparrow}(\mathbf{p})-w_{\downarrow}(\mathbf{p})}{w_{\uparrow}(\mathbf{p})+w_{\downarrow}(\mathbf{p})}$.

%\section{Results and discussion}

Figure \ref{fig1}(b) is a schematic of the formation of photoelectron energy spectrum. In details,
the ground state ($^2P_{3/2}$) absorbs the 11th and 13th harmonics,
resulting in the generation of two main peaks (H11 and H13). Meanwhile, the absorption and/or emission of
a IR photon from two neighboring main harmonic peaks generate a photoelectron interference SB12
between the two quantum pathways, serving as an interferometer for timing the photoelectron emission dynamics.
This process of two-photon transition is also suitable for the first excited state ($^2P_{1/2}$) of the core.
Consequently, the total photoelectron energy spectrum is
composed of two distinct groups of energy spectrum originating from the two ionic states, which are well-distinguished by the energy difference $\Delta I_p=0.66$ eV.

\begin{figure}[tb]
  \centering
  \includegraphics[scale=0.28]{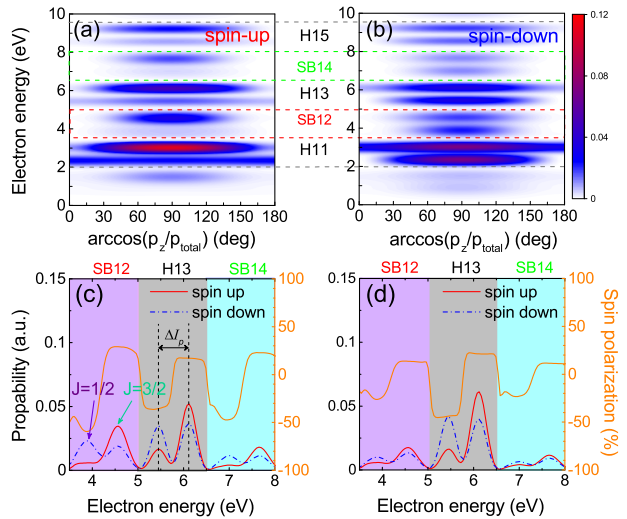}
  \caption{TDSE simulations of delay-integrated angle-resolved photoelectronenergy spectra for spin-up (a) and spin-down (b) electrons in the co-rotating geometries.  The photoelectron emission angle in the co-polarization plane, $\phi=\arctan(p_y/p_x)$, was integrated over its $2\pi$ range, and the XUV-IR delay was integrated over two IR optical-cycle periods. Photoelectron-energy spectra and the corresponding spin polarization in the light-polarization plane, that is, $\theta=\arccos(p_z/p_{\mathrm{total}})=90^\circ$ for
  the co-rotating (c) and counter-rotating (d) geometries. The  vertical dashed lines in (c) indicate the potential energy difference $\Delta I_p=0.66$ eV between $^2P_{1/2}$ state and $^2P_{3/2}$ state of the core.}\label{fig2}
\end{figure}

%**********************************************************************
Theoretically, according to Eq.~(\ref{up}), to obtain energy spectrum of spin-up and spin-down photoelectrons, we calculate the three-dimensional momentum distribution of the three degenerate orbitals from the outermost valence shell for two ionic states ($^2P_{3/2}$ and $^2P_{1/2}$).
For experiment \cite{Hartung2016,Trabert2018}, the spin-up and spin-down photoelectrons in the polarization plane are collected by a commercial Mott spin polarimeter \cite{Burnett1994} after traveling through a time-of-flight spectrometer.

Figs.~\ref{fig2}(a) and \ref{fig2}(b) show the calculated angle-resolved photoelectron energy spectrum of spin-up and spin-down photoelectrons, respectively, in the co-rotating case.
For both spin orientations, the total yield of three degenerate orbitals in the polarization plane dominates compared to that in other directions, which is analogous to the case of $p_+$ initial state, confirming the fact that the co-rotating electronic orbital ($m=1$) is preferentially removed in left circularly polarized XUV pulse.
We observe three MPs of the electron vortices corresponding to the
photoionization by H11, H13, and H15, as well as two SBs (SB12 and SB14).
The contributions from two ionization channels $^2P_{3/2}$ and $^2P_{1/2}$ are well separated in energy spectrum by a slight energy shift.
In the case of spin-up electrons, the double-peak structure in the MPs is clear, while the lower energy peak in the SB is strongly suppressed. However, for spin-down electrons, the two peaks in each SB are visible.
These phenomena are much more clearly seen in Figs.~\ref{fig2}(c) and \ref{fig2}(d) for the co-rotating and counter-rotating cases, respectively.
In the polarization plane, the contribution of the $^2P_{3/2}$ channel dominates for spin-up photoelectrons, while the yields from the $^2P_{1/2}$ channel are much weaker.
In contrast, for spin-down photoelectrons, the contributions from the $^2P_{3/2}$ and $^2P_{1/2}$ channels are comparable.
This is due to the fact that, for spin-up photoelectrons, the predominant yield of initial state $p_+$ is comprised of the yield from only the $J=3/2$ channel, according to Eq.~(\ref{up}).

In addition, the stronger suppression of $J=1/2$ channel in SB12 for spin-up photoelectrons indicates a higher spin polarization (reaches $63\%$ at approximately $E=3.8$ eV).
It is noteworthy that the energy-resolved spin polarization oscillates with increasing photoelectron energy.
In left circularly polarized XUV-APT, the SB photoelectrons from the $^2P_{3/2}$ channel have positive spin polarization, while those from the $^2P_{1/2}$ channel have negative spin polarization, regardless of whether
the IR field is in co-rotating or counter-rotating geometries [Figs.~\ref{fig2}(c) and \ref{fig2}(d)].
The calculated spin polarizations between two channels approximately follow the relation $S_{J=1/2}=-2S_{J=3/2}$, which is consistent with the measured results in \cite{Trabert2018}.
Notably, for SB12, the yield difference between spin-up and spin-down photoelectrons in co-rotating case is more apparent than that in counter-rotating case. This interesting phenomenon has not yet received much attention, or theoretical explanation.

\begin{figure}[tb]
  \centering
  \includegraphics[scale=0.1]{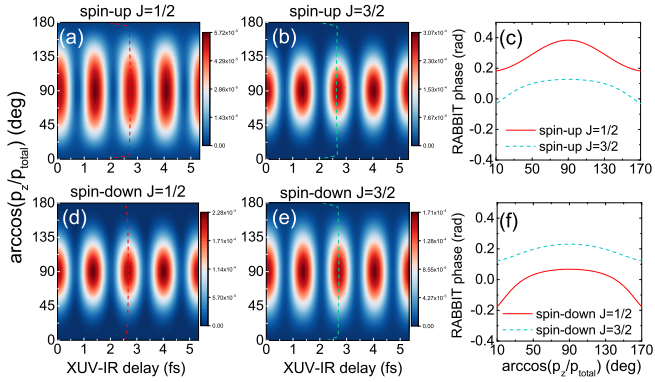}
  \caption{Delay- and angle-resolved photoelectron spectra of SB12 in the co-rotating geometry for spin-up electrons (upper row) and spin-down electrons (lower row) with ionic states $^2P_{1/2}$ (a,d) and $^2P_{3/2}$ (b,e). Extracted angle-resolved RABBIT phases from (a,b) and (d,e) are shown in (c) and (f), respectively. These corresponding RABBIT-phase curves are also plotted in (a,b) and (d,e) as dashed lines.}\label{fig3}
\end{figure}

Is there any correlation between the difference in spin-orientation-resolved ionization rates of SBs and the corresponding Wigner time delay for co-rotating and counter-rotating cases? To verify our hypothesis, we focus on the angle-resolved Wigner time from two channels of spin-up and spin-down photoelectrons.
In RABBIT, the angle-resolved SB yield oscillates with a period of $2\omega$ ($\omega$ is the IR center frequency) as the XUV-IR delay varies.
The phase of the yield oscillation, also called the RABBIT phase, contains the attochirp of the XUV field and the phases of two-photon transition amplitudes of photoelectrons.
In Fig.~\ref{fig3}, we represent the angle-resolved photoelectron spectra of SB12 as the XUV-IR delay varies in the co-rotating geometry [Figs.~\ref{fig3}(a), \ref{fig3}(b), \ref{fig3}(d), \ref{fig3}(e)], and the corresponding extracted angle-resolved RABBIT phases [Figs.~\ref{fig3}(c) and \ref{fig3}(f)].
Here, the photoelectron energy is integrated over $[3.43,4.26]$ eV
and $[4.26,5.04]$ eV for the $^2P_{1/2}$ and $^2P_{3/2}$ ionization channels, respectively, and the  photoelectron emission angle in the polarization plane $\varphi$ is fixed at $90^\circ$.
Our TDSE simulations demonstrate that the photoelectrons emitted from different angles with respect to the light polarization direction exhibit the different phases of the yield oscillation.

As illustrated in Figs.~\ref{fig3}(c) and \ref{fig3}(f),
we extract the angle-resolved RABBIT phase from the delay-resolved photoelectron angular distributions
by performing the Levenberg-Marquardt algorithm \cite{nocedal2006numerical}, for spin-up and spin-down photoelectrons, respectively.
These RABBIT-phase curves are also added to the 2D RABBIT trace plots by dashed lines for clarity.
In the co-rotating geometry, the angle-resolved RABBIT phases of spin-up and spin-down photoelectrons exhibit a downward bending characteristic with a similar curvature.
One distinct feature is that, for spin-up photoelectrons, the RABBIT phase from $^2P_{1/2}$ is always larger than that from $^2P_{3/2}$ channel, while the opposite is true for spin-down photoelectrons.
The SB time delay obtained from the RABBIT phase is given by $\tau_{\mathrm{SB}}=\phi_\mathrm{RABBIT}/2\omega=\tau_{\mathrm{atto}}+\tau_{\mathrm{atomic}}$ \cite{Paul2001,Cirelli2018},
where $\tau_{\mathrm{atto}}$ describes the group delay of the XUV-APT and is proportional to the phase difference between consecutive harmonics,
while $\tau_{\mathrm{atomic}}$ represents
the atomic scattering time delay due to the two-photon XUV+IR ionization process \cite{Klunder2011,Gong2022}.
Since the angle-resolved RABBIT phase in this work is only considered in one SB, the influence of XUV attochirp on the RABBIT phase, which only depends on the photon energy, can be excluded.
Therefore, the spin-up photoelectrons emitted from $^2P_{1/2}$ channel are delayed in time compared to those emitted from $^2P_{3/2}$ channel, while the spin-down photoelectrons emitted from $^2P_{1/2}$ are advanced temporally compared to those
emitted from $^2P_{3/2}$ channel.

\begin{figure}[tb]
  \centering
  \includegraphics[scale=0.2]{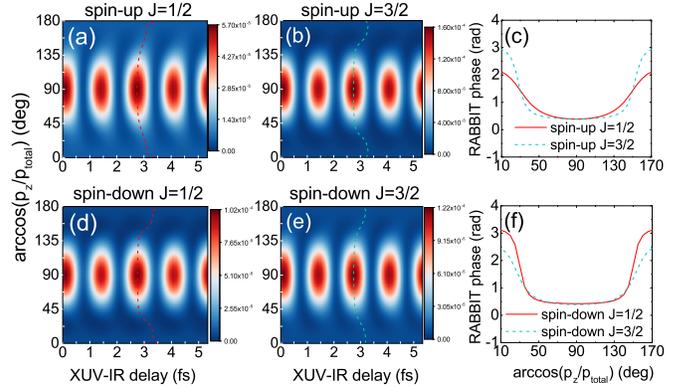}
  \caption{The same as Fig.~\ref{fig3} but in the counter-rotating geometry.}\label{fig4}
\end{figure}

We observe a significant difference in the angle-resolved RABBIT phase between the co-rotating and counter-rotating geometries. As depicted in Fig.~\ref{fig4}, the dichroic structures in the counter-rotating geometry exhibit opposite bending directions and have enhanced curvatures compared to those in the co-rotating geometry.
We also find that the RABBIT phase is highly sensitive to the emission angle when the photoelectrons emitted close to the light propagation direction. However, for the photoelectrons emitted close to the light-polarization plane, the RABBIT phase responses dully to the variation of emission angle.
Furthermore, the spin-up photoelectrons from $^2P_{1/2}$ and $^2P_{3/2}$ channels differ considerably in the RABBIT phase when the emission angle $\theta=0^\circ$ or $180^\circ$, while this difference decreases significantly around emission angle $\theta=90^\circ$ [Fig.~\ref{fig4}(c)], and the same is true for spin-down photoelectrons [Fig.~\ref{fig4}(f)].

\begin{figure}[tb]
  \centering
  \includegraphics[scale=0.17]{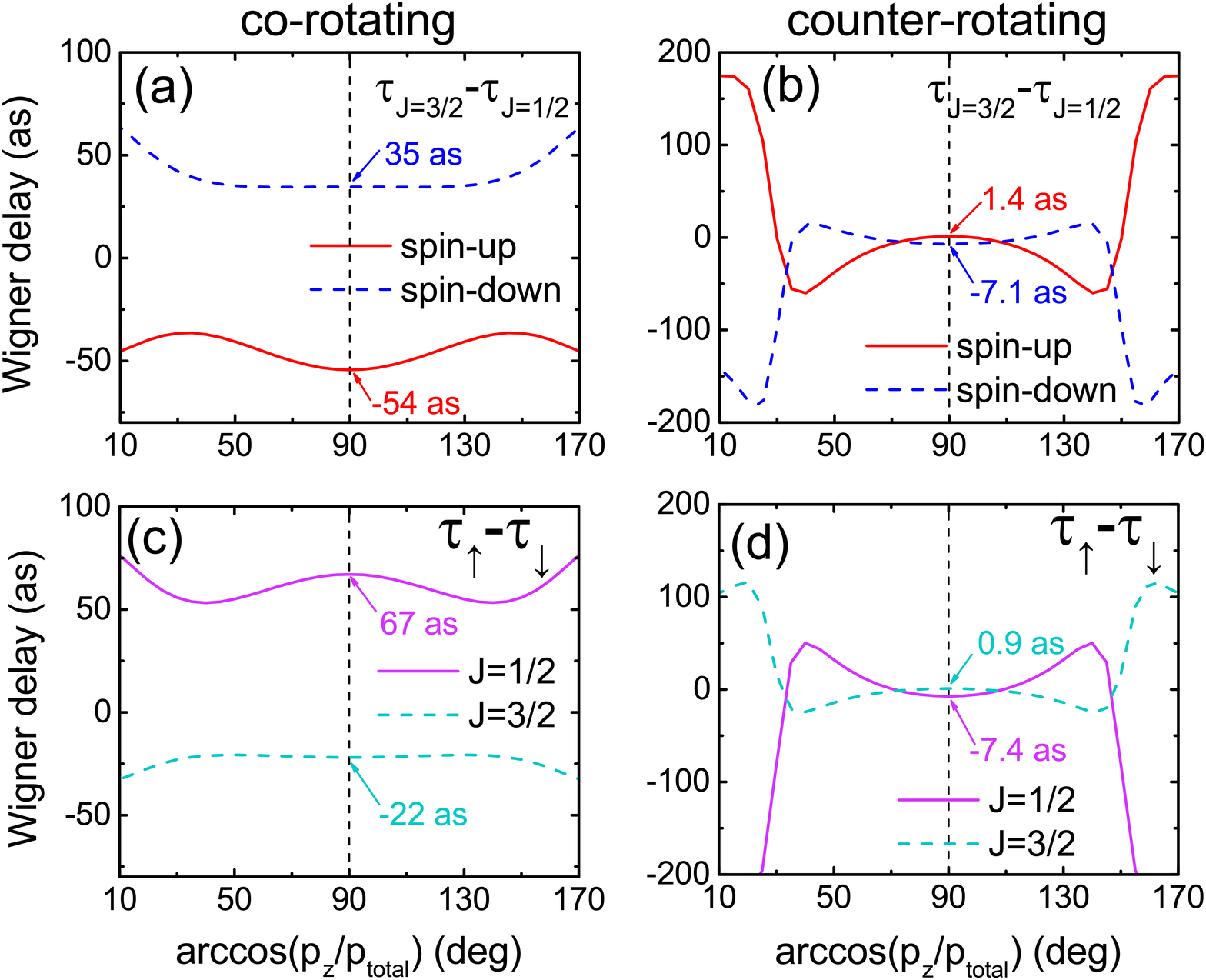}
  \caption{Wigner time delay between $^2P_{3/2}$ and $^2P_{1/2}$ channels of spin-up and spin-down electrons in the co-rotating (a) and counter-rotating (b) geometries as a function of the emission angle $\theta$. Also shown is the Wigner time delay between spin-up and spin-down electrons for two ionization channels in the co-rotating (c) and counter-rotating (d) geometries. The colorful arrows indicate the corresponding Wigner delay in the polarization plane.}\label{fig5}
\end{figure}

The two-photon transition-induced atomic photoionization time delay $\tau_{\mathrm{atomic}}$ can be decomposed into two distinct components: $\tau_{\mathrm{atomic}}=\tau_\mathrm{W}+\tau_{cc}$.
The former represents the Wigner delay of one-photon ionization,
while the latter is a continuum–continuum delay induced by the electron being probed by an IR
laser field in a long-range potential with a Coulomb tail, as previously demonstrated in \cite{Dahlstroem2012jpb}.
Importantly, we note that $\tau_{cc}$ depends solely on the frequency of the IR probe field, the Coulomb field of the parent ion, and the final kinetic energy of the released photoelectron, as discussed in previous works \cite{Zhang2010,Dahlstrom2012,Dahlstroem2013}.
Using this information, we are able to extract the spin-orientation- and ionization-channel-resolved Wigner time delay from the RABBIT phase, as shown in Fig.~\ref{fig5}.
Our analysis reveal interesting differences between the co-rotating and counter-rotating geometries.
Specifically, for the co-rotating case, the Wigner time delays between two channels of spin-up and spin-down photoelectrons
have opposite signs and are clearly distinguishable as the emission angle varies [Fig.~\ref{fig5}(a)].
For instance, in the polarization plane, the Wigner delay is 35 as and $-54$ as for spin-up and spin-down photoelectrons, respectively.
On the other hand, for the counter-rotating case, the Wigner time delay changes drastically, making it difficult to distinguish the spin direction of the photoelectrons in the polarization plane with a mere few attoseconds difference, as shown in Fig.~\ref{fig5}(b).
Further, we extract the spin-orientation-resolved Wigner time delay from $^2P_{1/2}$ and $^2P_{3/2}$ channels in the co-rotating [Fig.~\ref{fig5}(c)] and counter-rotating cases [Fig.~\ref{fig5}(d)].
Our results show that in the co-rotating geometry, spin-up photoelectrons in the polarization plane emitted from the $^2P_{1/2}$ ionization channel are delayed by 67 attoseconds compared to spin-down photoelectrons emitted from the same channel.
In contrast, for the $^2P_{3/2}$ channel, spin-up photoelectrons are advanced by 22 attoseconds.
Notably, in the counter-rotating case, the Wigner time delays of the two channels are challenging to differentiate by spin orientation.

%\section{Conclusions}
In summary, we have investigated the retrieve of the spin-orientation-resolved Wigner time delay in strong field ionization of Kr atoms using the circular RABBIT technique for the first time.
This investigation presented the first successful attempt at achieving this feat, and we achieve it by leveraging the fact that
the spin-resolved ionization rates of photoelectrons emitted from $^2P_{1/2}$ and $^2P_{3/2}$ channels can be expressed
via $m$-resolved ionization rates of initial state.
We observe a significant yield suppression of spin-up photoelectrons from $^2P_{1/2}$ channel at the SBs, while an enhancement at the same SB from $^2P_{3/2}$ channel, as compared to the spin-down photoelectrons in the co-rotating geometry.
In contrast, for the counter-rotating geometry, we observe comparable yields between spin-up and spin-down photoelectrons.
Importantly, the yield difference in the polarization plane encods the different behaviors of the Wigner time delay in the two geometries. Lastly, we demonstrate that the extracted Wigner time delay from the RABBIT phase could be well-resolved by spin orientation in the co-rotating geometry but not in the counter-rotating geometry.
Our approach opens up a new avenue for probing the spin-orientation-resolved Wigner time to achieve
the attosecond chronoscopy and exploring the photoelectron spin dynamics in the noble gas atoms. We are optimistic that these results will spur experimental measurements of this exciting spin-orientation-resolved attosecond chronoscopy method.

%\section*{Acknowledgements}
This work is supported by the National Natural Science Foundation of China (NSFC) (Grant Nos. 12074265, 12204314, 12147117, and 61775146),
Guangdong Basic and Applied Basic Research Foundation (Grant No. 2022A1515010329).

%%%%%%%%%% If preparing manually:
\bibliography{rabbit}

\end{document}